\documentclass[sigconf]{acmart}

\AtBeginDocument{%
  \providecommand\BibTeX{{%
    \normalfont B\kern-0.5em{\scshape i\kern-0.25em b}\kern-0.8em\TeX}}}

\copyrightyear{2020} 
\acmYear{2020} 
\setcopyright{acmcopyright}
\acmConference[SIGIR '20]{Proceedings of the 43rd International ACM SIGIR Conference on Research and Development in Information Retrieval}{July 25--30, 2020}{Virtual Event, China}
\acmBooktitle{Proceedings of the 43rd International ACM SIGIR Conference on Research and Development in Information Retrieval (SIGIR '20), July 25--30, 2020, Virtual Event, China}
\acmPrice{15.00}
\acmDOI{10.1145/3397271.3401403}
\acmISBN{978-1-4503-8016-4/20/07}

\usepackage{adjustbox}
\usepackage{xcolor}
\newcommand{\hl}[1]{#1}
\begin{document}
% Disable headers
\fancyhead{}
%%
%% The "title" command has an optional parameter,
%% allowing the author to define a "short title" to be used in page headers.
\title{Deep Job Understanding at LinkedIn}

%%
%% The "author" command and its associated commands are used to define
%% the authors and their affiliations.
%% Of note is the shared affiliation of the first two authors, and the
%% "authornote" and "authornotemark" commands
%% used to denote shared contribution to the research.

\author{Shan Li \quad Baoxu Shi \quad Jaewon Yang \quad Ji Yan \quad Shuai Wang \quad Fei Chen \quad Qi He}
\affiliation{\institution{LinkedIn}}
\email{{shali,dashi,jeyang,jyan,shuawang,fechen,qhe}@linkedin.com}

%%
%% By default, the full list of authors will be used in the page
%% headers. Often, this list is too long, and will overlap
%% other information printed in the page headers. This command allows
%% the author to define a more concise list
%% of authors' names for this purpose.
\renewcommand{\shortauthors}{Li et. al.}

%%
%% The abstract is a short summary of the work to be presented in the
%% article.
\begin{abstract}
As the world’s largest professional network, LinkedIn wants to create economic opportunity for everyone in the global workforce. One of its most critical missions is matching jobs with processionals. Improving job targeting accuracy and hire efficiency align with LinkedIn's Member First Motto. To achieve those goals, we need to understand unstructured job postings with noisy information. We applied deep transfer learning to create domain-specific job understanding models. After this, \hl{jobs are represented by professional entities}, including titles, skills, companies, and assessment questions. To continuously improve LinkedIn’s job understanding ability, we designed an expert feedback loop where we integrated job understanding models into LinkedIn’s products to collect job posters’ feedback. In this demonstration, we present LinkedIn’s job posting flow and demonstrate how the integrated deep job understanding work improves job posters’ satisfaction and provides significant metric lifts in LinkedIn's job recommendation system.
\end{abstract}

%%
%% The code below is generated by the tool at http://dl.acm.org/ccs.cfm.
%% Please copy and paste the code instead of the example below.
%%

%%
%% Keywords. The author(s) should pick words that accurately describe
%% the work being presented. Separate the keywords with commas.

%%
%% This command processes the author and affiliation and title
%% information and builds the first part of the formatted document.
\maketitle

\section{Introduction}
LinkedIn serves as a job marketplace that matches millions of jobs to more than 675 million members. To create economic opportunity for every member of the global workforce, LinkedIn needs to understand the job marketplace precisely. However, understanding job postings is non-trivial due to its lengthy and noisy nature. Job postings usually cover a wide range of topics ranging from company description, job qualifications, benefits to disclaimers. It is challenging to model job postings directly in tasks such as job recommendation and applicant evaluation. To address this challenge, we develop job understanding models that take noisy job postings as input and output structured data for easy interpretation. To be specific, \hl{we standardize job postings into professional entities} that represent the characteristics of a job, for example, \hl{the occupation of a job (title), the hiring company (company), key skills required by the job (skills), and job qualifications (assessment questions)}.

Standardizing job information has a tremendous impact on the LinkedIn ecosystem. Firstly, it helps recruiters to do better candidate targeting. Using the extracted key skill entities, recruiters can target the candidates that have the right skill set. Secondly, it helps members to find jobs easily. It is convenient for members to search jobs based on the professional entities standardized from the job postings such as occupation and requirements. Lastly, it improves the overall hire efficiency. By extracting assessment questions and key skill entities from jobs, LinkedIn can automatically evaluate job-applicant fit by comparing them with member-side entities.

However, developing a good job understanding model is a challenging task in many aspects. 
First, we need to define an extensive professional entity taxonomy that covers a wide range of industries and occupations. Without a comprehensive taxonomy, it is hard to represent jobs using entities.
Second, we need domain-specific natural language understanding models to understand jobs. Compared to ordinary articles, job posting text is often long, noisy, and has job-specific writing styles, general-purpose Natural Language Processing (NLP) models are less suitable for this task. 
Third, job understanding models need to be market-aware. To be specific, it needs to go beyond simply identifying mentioned entities and understand market importance of entities via modeling each different job market and the hiring experts in the market.

Unfortunately, existing methods haven’t addressed all the above challenges. Models such as SPTM~\cite{xu2018measuring} and TATF~\cite{wu2019trend} perform job-skill analysis on IT skills only. DuerQuiz~\cite{qin2019duerquiz} focuses on job content only and ignores the market dynamics. Lastly, none of these models explicitly model market variance via establishing a feedback loop between models and hiring experts.

In this work, we present LinkedIn's deep job understanding and demonstrate LinkedIn's job posting flow\footnote{\url{https://www.linkedin.com/talent/job-posting/post}} powered by our work. 
We combined machine learning techniques and linguist experts to curate the world's largest professional entity taxonomy. 
To develop domain-specific content understanding model, we used deep transfer learning to adapt open-domain NLP models and professional entity embeddings trained on LinkedIn member profiles to our domain. 
We engineered market-specific features and established a feedback loop with job posters. We showed the model outputs and allowed them to override the suggestions to collect the feedback. By developing the feedback loop, we were able to collect domain-specific and market-aware training data to continuously improve our model. Moreover, such a feedback loop also empowered the job posters while giving options to them to delegate decisions to AI models. \hl{More importantly, the standardized job data improves member experience in several downstream LinkedIn products such as Job Search, Job Alert and Job You Maybe Interested In (JYMBII)}~\cite{kenthapadi2017personalized}.

\section{Related Work}
Named Entity Recognition (NER) is most related to our work given the present knowledge. However, there are a few key distinctions between general NER and our work. First, most neural network (NN) NER models~\cite{akbik2018contextual,peters2018deep,gupta2018deep} are trained on open domain corpora~\cite{sang2003introduction,weischedel2013ontonotes} and focus on recognizing a limited set of entities including person, location, date, and organization. These methods are not designed to recognize professional entities. In stark contrast, few work is focused on recognizing professional entities such as title and skills in job recruiting fields. Goindani, et al.~\cite{goindani2017employer} recognize industries from job postings by treating it as a classification problem. Qin, et. al.~\cite{qin2019duerquiz} developed a NN model to extract skill entities from the job postings and resumes. Yet, they solely focus on one specific entity extractor, neglecting mutual benefits from other entity extractors. Additionally, there is no feedback loop~\cite{cruz2012interactive,yan2019social} in their development cycle. This is a major shortcoming because models based on their approaches fail to adapt well with job market fluctuations and changes. To our best knowledge, as an important part of our work, feedback loop has not been applied to the entities standardization problem in the job domain yet.

\section{Job Standardization}

\subsection{Architecture}
\begin{figure}[t] 
\centering
    \includegraphics[width=\linewidth]{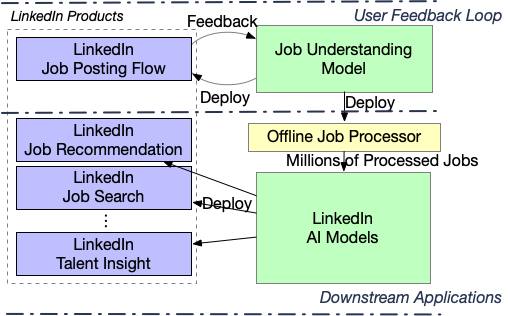}
    \caption{Overview architecture of the user feedback loop and LinkedIn's ecosystem around job understanding.}
\label{fig:architecture} 
\end{figure}

Fig.\ref{fig:architecture} shows \hl{the overall architecture of the job standardization and how it fits in} LinkedIn’s ecosystem. It essentially consists of 2 phases: user feedback loop and downstream applications. In the user feedback loop, we first initialize our work with small-scale human annotated data targeting a specific job standardization task. With the small amount of high quality data and our professional entity taxonomy, we are able to train a simple linear or tree-based job understanding model as our bootstrapping solution and deploy it to the production. The bootstrapping model also starts feeding freshly standardized data to other downstream AI models to play with. As the bootstrapping model serves the customer online, more user-behavior and market-specific data will be collected. A more advanced NN-based job understanding model is trained and deployed to replace the bootstrapping model. \hl{And it will loop once a few months to constantly improve and adjust the model with more latest online data being collected.} \hl{Once trained, the latest job standardization models will be deployed to a nearline streaming platform to process millions of LinkedIn's jobs and the standardized data will further be fed to all downstream AI models for consumption at LinkedIn within minutes of latency}. The downstream products consume the data and improve the model before being re-deployed for 
online serving. The collection of LinkedIn AI products forms an ecosystem around the job standardization work and keeps benefiting users and serving the job market.

\subsection{Models}
We developed a set of job standardization models to identify professional entities from job postings, and used an user feedback loop to continuously improve model performance by retraining it with collected data. In general, We developed these models using a two-step procedure: entity tagging (i.e. candidate generation) and content \& market-aware entity ranking. To achieve the best performance, we trained type-specific models for each entity. Currently we support four types of entities: title, skill, company and assessment question.

\subsubsection{Entity Tagging}
We built one of the world’s largest in-house professional entity taxonomy with 30k titles and 50k skills by our taxonomy linguists and machine learning models. Utilizing the comprehensive taxonomy with rich entity aliases, we developed string-based entity taggers for each entity type. We generated entity candidate sets by identifying all possible entity mentions from the job postings and then pass them to the entity ranking model.

\subsubsection{Content \& Market-aware Entity Ranking}
We designed deep NN models to rank the candidates and pick the top-k most important standardized entities in a given job posting. In order to get the mastery of both entities and context, we built rich feature set to represent the content. To capture shallow linguistic structure and meanings of the text, we adopt engineered features such as job location, email domain, n-gram matching distance, etc. To extract deep semantic meaning of the entities and context, we use deep transfer learning to adapt pre-trained NLP models such as Deep Averaging Network~\cite{iyyer2015deep} and FastText~\cite{joulin2016bag} to the job domain, and apply them to model sentences and paragraphs that contain entity mentions. To model cross-tagger entity-entity relationships, we utilize entity embeddings learned from member profiles~\cite{shi2019representation}. We fine-tune those embeddings to measure entity coherence in jobs and detect potential outliers among the entity candidates.

We also rank the candidates based on market-related factors to do personalized recommendations for different markets, industries, and regions. We do this in two ways. First, we construct member-job Pointwise Mutual Information (PMI) features for all jobs and members to have an grip on the overall personnel-recruiting market. Also, by collecting the hiring expert signals such as users’ acceptance/rejection rate in the user feedback loop, we are able to master the latest market trends and iteratively refine our models.

\subsection{User Feedback loop}
We build a user feedback loop to continuously improve the job standardization. Specifically, we track and collect user feedback directly from recruiters and members from all industries. Therefore, we keep getting domain-specific and market-aware data that has better quality than the data labeled by human annotators who are generally not experts in specific hiring markets. Also, with the data pipeline automatically consuming the large-scale feedback data in a daily fashion, our model is highly scalable. By building a feedback loop, we use it for getting a good sense of job and hiring market and being sensitive to user behaviour changes. For example, It is used to calculate market-aware features that reveal the latest skill trend and most popular questions being asked in a specific industry.

\section{Job Posting Flow Demonstration}

In this section, we will showcase how the deep job understanding empowers LinkedIn's job posting flow. By applying the deep job understanding models and establishing feedback loops, we observed +11 Net Promoter Score (NPS) improvement indicating good user satisfaction, and $30\%$ more job applications because of the better job targeting that aims at putting the job post in front of the right candidates. Additionally, we observed the latest job skill standardization model reduces the need of manually adding job targeting skills by $33.75\%$ compared to the original model trained without user feedback signals. Also, $80\%$ of jobs with assessment questions get a qualified applicant within 24 hours. Lastly, with LinkedIn's jobs represented by professional entities, we see $+1.92\%$ onsite job apply and $+2.66\%$ job save in the job recommendation system.

\subsection{Title Standardization}

\begin{figure}[t] 
\centering
    \begin{adjustbox}{max width=.95\linewidth}
    \includegraphics[width=\textwidth]{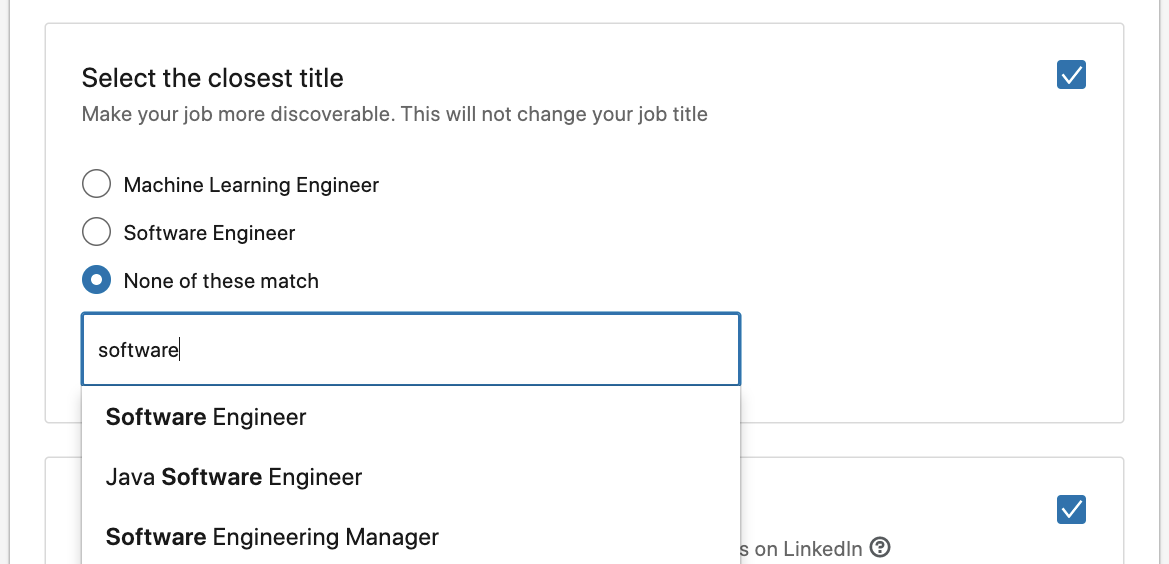}
    \end{adjustbox}
    \caption{Example of LinkedIn's title standardization model provides standardized title suggestions to job posters given job posting name ``Machine Learning Software Engineer''.}
\label{fig:j2t_demo} 
\end{figure}

When a recruiter posts a job on the LinkedIn platform, we standardize the job posting's title into a standardized title entity for easy index. The job title standardization consists of two steps. Firstly, the original title given by the user is broken into individual tokens to form a key lookup table for candidate retrieval. Secondly, the candidates are passed into a ranking model that outputs a confidence score. LinkedIn shows these standardized titles to the user to make a final selection. As shown in Fig.~\ref{fig:j2t_demo}, the user has the option of choosing a standardized title suggested by the model, or manually inputting any title as the user sees fit. In the latter case, LinkedIn provides a typeahead to encourage the user to input a standardized title in the end. By tracking the user's final choice of the title at the end of the job posting, we can use the feedback to train better models over time.

\subsection{Skill Standardization}
\begin{figure}[t] 
\centering
    \begin{adjustbox}{max width=.95\linewidth}
    \includegraphics[width=\textwidth]{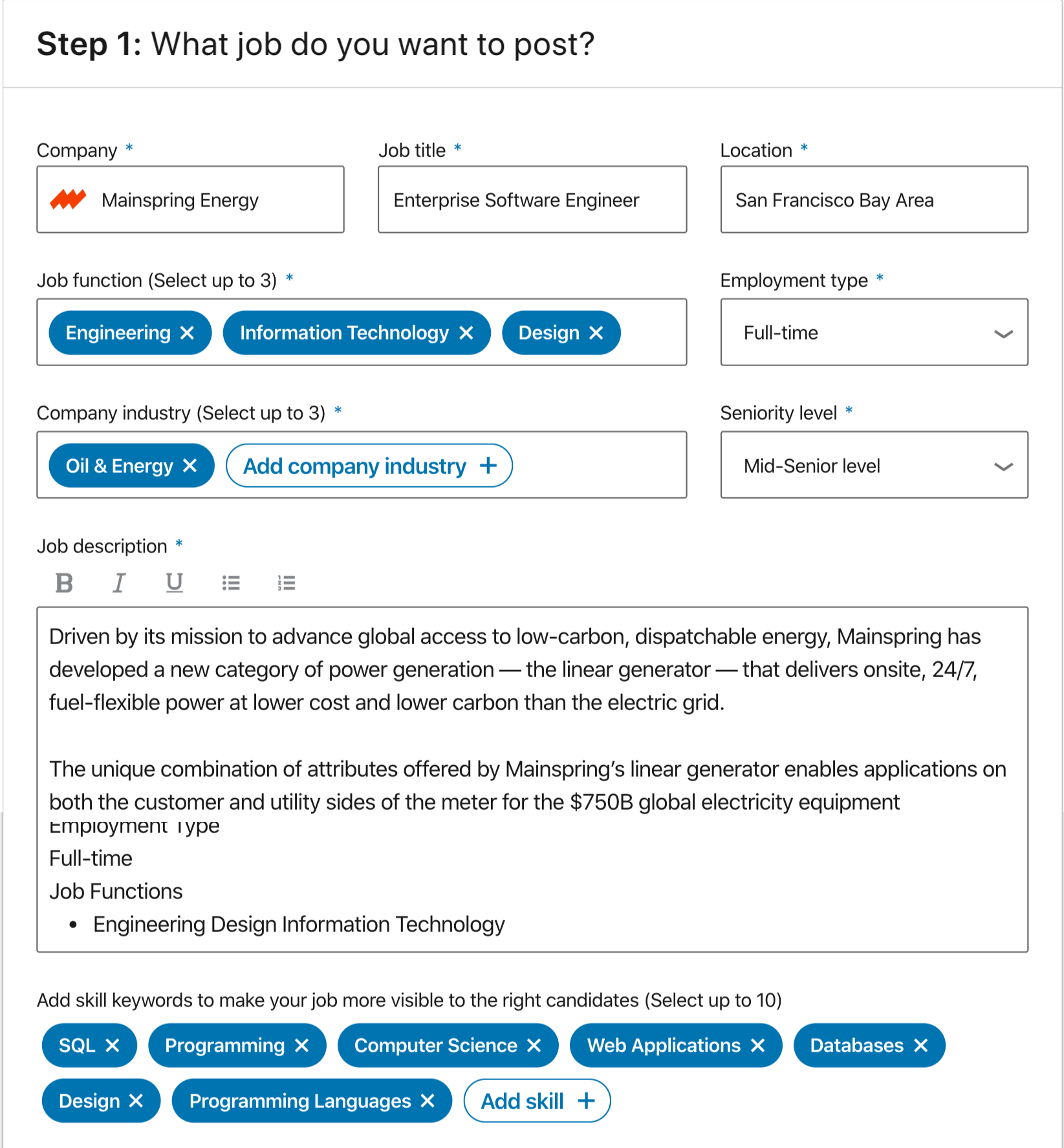}
    \end{adjustbox}
    \caption{LinkedIn's skill standardization model provides job targeting skill suggestions to job posters.}
\label{fig:j2s_demo} 
\end{figure}

At LinkedIn, we use a job skill standardizer to suggest job targeting skills for job posters. Fig.~\ref{fig:j2s_demo} shows an example on how job posters receive skill suggestions. When a recruiter fills in the basic information of a job, we will process the text inputs and provide job targeting skill suggestions to them. The skill standardizer will first generate a list of mentioned skills using the skill entity tagger, and then pass it into a learning-to-rank model to pick the most important job targeting skills based on a variety of content and market signals. For example, the model uses FastText to model the semantic meaning of job and uses entity embeddings to model the skill coherence within the job. It also uses signals aggregated from hiring experts to model the market trend.

By providing job posters with a set of skills which the job targets on instead of asking them to provide it manually, we improve the user satisfaction by reducing their workload. Moreover, because job posters still have the chance to reject or propose other job targeting skills, we are able to collect their feedback and hence refine our model based on their latest preferences.

\subsection{Company Standardization}
\begin{figure}[t] 
\centering
    \begin{adjustbox}{max width=.9\linewidth}
    \includegraphics[width=\textwidth]{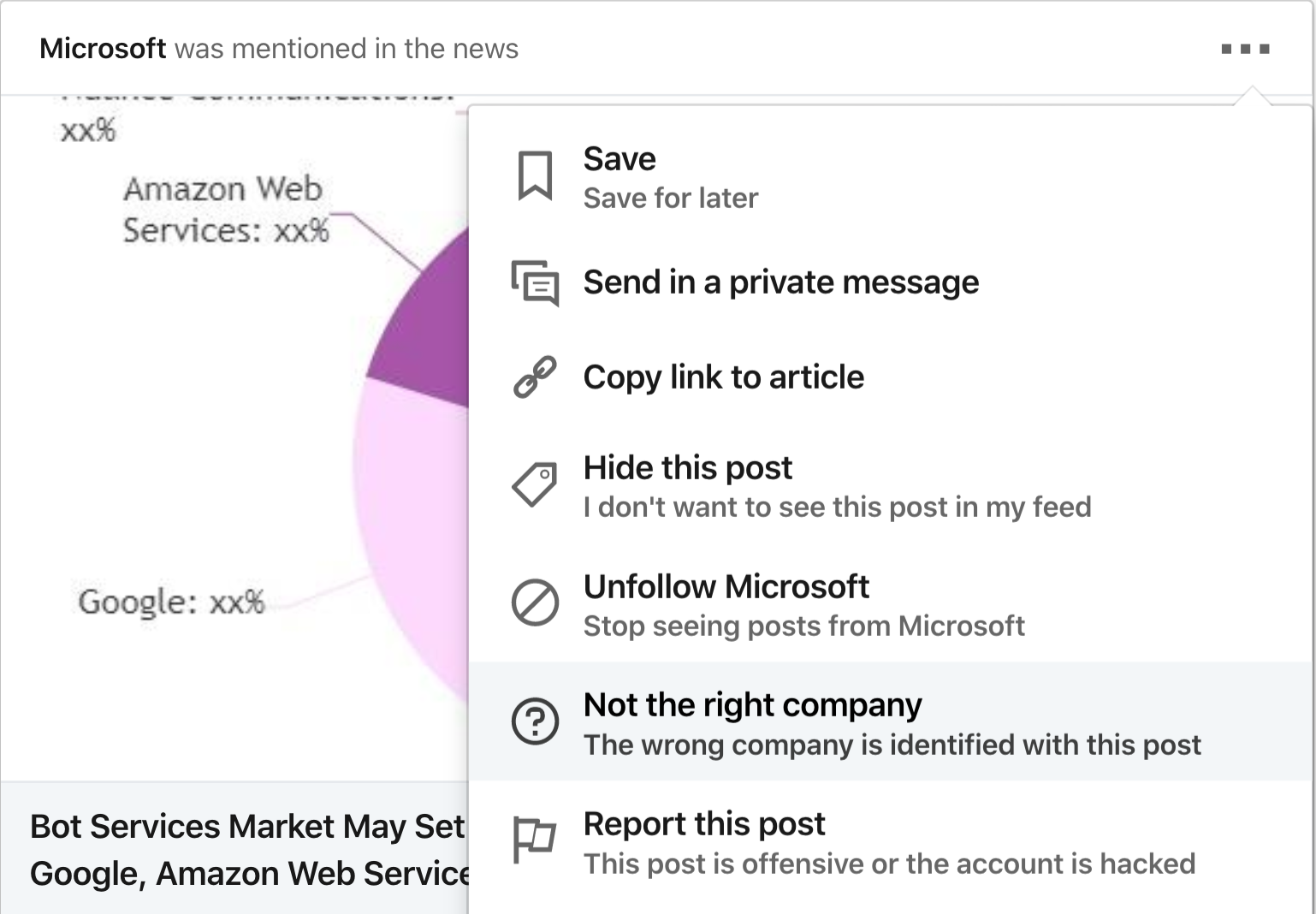}
    \end{adjustbox}
    \caption{Example of collecting company standardization data from LinkedIn members in the feedback loop.}
\label{fig:j2c_demo} 
\end{figure}

When a company is mentioned in news articles in LinkedIn feed, it is passed through our company standardizer to output a standardized company entity in our company taxonomy. Again, we establish an open channel here to take in user feedback. When the user selects “Not the right company” from the list as shown in Fig.~\ref{fig:j2c_demo}, we get chance to obtain indicative negative labels in the user feedback loop to further benefit the model.

\subsection{Assessment Question Standardization}
Lastly, we show how LinkedIn provides assessment question suggestions and collect feedback from job posters. The assessment question standardization model takes the job description provided by the job poster as input and outputs a list of important assessment questions extracted from the job posting~\cite{shi2020learning}. First, we apply a fine-tuned Deep Averaging Network model to the sentences in the job posting to generate a list of assessment question candidates. Next, we aggregate questions of sentences in job posting and build a rich set of features. We curate features not only from the text itself, but also from statistics captured in the whole job market such as Pointwise Mutual Information (PMI) scores between job industries and question types. Finally, a gradient-boosted tree-based model~\cite{chen2016xgboost} ranks all questions and returns the top k questions to the user. 

\begin{figure}[t] 
\centering
    \begin{adjustbox}{max width=.95\linewidth}
    \includegraphics[width=\textwidth]{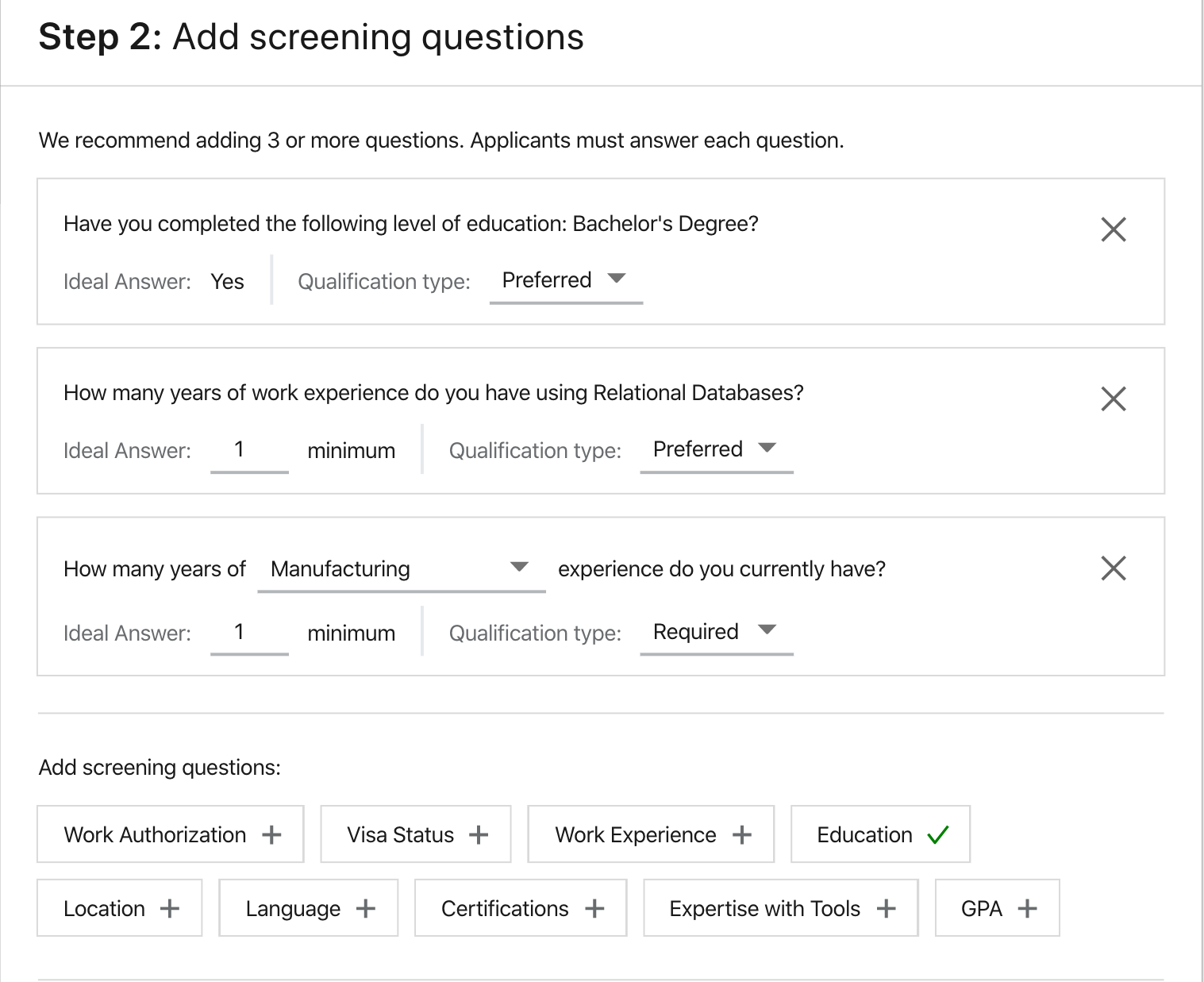}
    \end{adjustbox}
    \caption{LinkedIn's assessment question model provides screening question suggestions to job posters.}
\label{fig:j2q_demo} 
\end{figure}

As is shown in Fig.~\ref{fig:j2q_demo}, it suggests top three questions and returns them to the users, where they have options to choose to either accept or reject the suggested questions. By tracking user actions on the page, we collect the data that reflects the latest market trends and user preference. The user feedback loop forms when we iteratively update our model using the up-to-date feedback data and put it into the production to serve again.

\section{Conclusion}
At LinkedIn, we performed job standardization work over millions of jobs. We built the world’s largest and most complete professional entity taxonomy to categorize and standardize the professional entities in the job posting. We used deep transfer learning to develop title, skill, company, and assessment question entity taggers which standardize entities out of job postings. The models were tailored to be domain-specific and market-oriented, thereby keeping great sensitivity to different job markets. More importantly, we built and presented our customer feedback loop by tracking users’ interactive behaviour on current models in LinkedIn’s job posting flow. Using the feedback loop, we were able to iteratively adjust our models to let them not only be more powerful but also stay more sensitive to the job market changes. Our online A/B test results showed that these user feedback loops improved job posters' satisfaction, increased the performance of our job understanding models, and ultimately led to better matches between jobs and our members.
%%
%% The acknowledgments section is defined using the "acks" environment
%% (and NOT an unnumbered section). This ensures the proper
%% identification of the section in the article metadata, and the
%% consistent spelling of the heading.
% \begin{acks}
% We want to thank Feishe Chen, Shuai Wang, Ji Yan, and Peide Zhong for developing the models and conducting online experiments. We would like to than Fei Chen for her valuable feedback.
% \end{acks}

%%
%% The next two lines define the bibliography style to be used, and
%% the bibliography file.
\bibliographystyle{ACM-Reference-Format}
\bibliography{acmart}

\end{document}